\documentclass[pra,twocolumn,nofootinbib,preprintnumbers,superscriptaddress]{revtex4-1}

\usepackage{natbib}
\usepackage{amssymb,amsbsy,amsmath,amsfonts}
\usepackage{mathrsfs}
\usepackage{graphicx}% Include figure files
\usepackage{epsf,epsfig,float,latexsym,amsthm,fancyhdr,rotating}
\usepackage{graphics,psfrag,longtable}
\usepackage{slashed}
\usepackage{esint}
\usepackage{nicefrac}
\usepackage{braket}
\usepackage{mathtools}
\usepackage{ulem}

\newcommand{\nn}{\nonumber}
\renewcommand{\Im}{{\rm Im\,}} 

\renewcommand{\eth}{\partial}
\newcommand{\be}{\begin{equation}}
\newcommand{\ee}{\end{equation}}

\usepackage{color}
\definecolor{purple}{rgb}{.36,.12,.60}
\definecolor{orange}{rgb}{.9,.3,.0}
\definecolor{green}{rgb}{0,.6,0}

\usepackage[colorlinks,citecolor=blue,linktoc=all,linkcolor=cyan]{hyperref}

\begin{document}

\title {Vorticity of Twisted Spinor Fields}

\author{Andrei Afanasev}

\affiliation{Department of Physics,
The George Washington University, Washington, DC 20052, USA}

\author{Carl E. Carlson}

\affiliation{Physics Department, William \& Mary, Williamsburg, Virginia 23187, USA}

%\affiliation{And ... ?}

\author{Asmita Mukherjee}

\affiliation{Department of Physics, Indian Institute of Technology Bombay, Powai, Mumbai 400076, India}

\begin{abstract}
Spinor fields with a vortex structure in free space that allow them to have arbitrary integer orbital angular momentum along the direction of motion have been studied for some time.  Relatively new is the observation in a certain context that the vortex center of this field structure is, unlike a classical whirlpool, not singular.  We point out that there are several ways to calculate the local velocity of the spinor field and that all but one show a singular vorticity at the vortex line.  That one, using the Dirac bilinear current with no derivatives, is the only one so far (to our knowledge) studied in the literature in this context and we further show how to understand an apparent conflict in the existing results.
\end{abstract}
\date{\today
}
\maketitle

%%%%%%%%%%%%%%%%%%%%%%%%%%%%%%%%%%

\section{Introduction}

%%%%%%%%%%%%%%%%%%%%%%%%%%%%%%%%%%
Vorticity is given by the curl of the velocity of a spinor field at a given space-time point. Two overlapping questions in the context of spinor fields with non-uniform wavefronts propagating in free space are how singular the vortex lines may be and what expressions should be used to calculate the momentum densities of structured spinor, here mainly spin 1/2, fields.   The discussion can be phrased in terms of twisted spinor states, which are wave packets with a definite overall direction of motion and which can have arbitrary integer, times $\hbar$, orbital angular momentum along the propagation direction; reviews can be found in, for example, Refs.~\cite{2017PhR...690....1B,2017RvMP...89c5004L,2018ConPh..59..126L,2022arXiv220500412I}. % The discussion can be phrased in terms of electrons states, but of course extends to twisted states of any spin-1/2 field.  
The twisted spinor state has local momentum densities with azimuthal components that swirl about a vortex line whose direction is set by the overall propagation direction.  

One question is whether the state has a true vortex, in the sense that the local momenta or local field velocities have a singular curl at the location of the vortex line.  Quickly come questions of how one should calculate the local momenta or momentum densities that give the vorticity.  We may list three possibilities for the momentum densities.  One follows from the Dirac equation, particularly when there are electromagnetic interactions, where one can define a bilinear conserved current.
%, {\color{blue} using no derivatives (this phrase is not needed)}
 Momenta and thence velocities follow, analogously to electromagnetic currents of classical particles, as ratios of the spatial and temporal components of the corresponding four-vectors.  We refer to this as obtaining the momentum density field or the velocity field from the Dirac current.  Alternatively, in field theory,  one can start from either the canonical or the symmetric Belinfante-Rosenfeld version of the energy-momentum tensor to obtain the momentum density.  For a plane wave state, or a momentum eigenstate, the three ways to obtain a momentum density give the same results,  however in a general case, one gets three different numerical answers.

 Two papers that introduce the question about the singularity of the vortex line are by Bialynicki-Birula and Bialynicka-Birula~\cite{2017PhRvL.118k4801B} and by Barnett~\cite{2017PhRvL.118k4802B}, with further commentary in~\cite{2017PhyOJ..10...26L,2017PhRvL.119b9501B,2017PhRvL.119b9502B} and more recently in~\cite{2020NatSR..10.7417H}. These papers each give coordinate space solutions for twisted electrons valid, in particular, at small distances from the vortex line.  They then analyze the vorticity of the velocity field, in both of these papers obtaining the momentum density and velocity field from the Dirac current definition.  They give, interestingly enough, different opinions on whether a true vortex exists.

The papers just mentioned~\cite{2017PhRvL.118k4801B,2017PhRvL.118k4802B} find solutions by unique and interesting methods, and are not the same in appearance.  There is also an older way~\cite{2011PhRvL.107q4802B,2017PhR...690....1B} to obtain the twisted spinor fields, from knowledge of twisted states in momentum space
We will begin by reviewing the solutions, and will observe that the solutions are the same in the sense that all can be expressed as linear combinations of each other.   

The different conclusions are a matter of interpretation.  Strictly speaking, as we will show in Sec.~\ref{sec:two}, when getting the velocity field from the Dirac current the vortex line does not have a singular vorticity. A classical vortex, and we will define our use of that term below, has singular vorticity at the vortex center.  However, as particularly noted by Ref.~\cite{2017PhRvL.118k4802B}, in a non-relativistic or moderately relativistic situation,  there are solutions where the twisted electron velocity field is like a classical vortex down to distances of order of an electron Compton wavelength, about 2.4 picometers, from the vortex line.  This is a very small distance on an atomic scale, and leads Ref.~\cite{2017PhRvL.118k4802B} to conclude that for practical purposes, the twisted electron behaves (or at least can behave) like a state with a singular vortex line. Ref.~\cite{2017PhRvL.118k4801B} shows two solutions, one of which finds no classical vortex even at larger distances, and both of which find no singularity at the vortex line.  Their conclusion is the strict one that the vorticity of the vortex line is never singular. 

These conclusions involving different opinions were in the context of a momentum density or velocity field obtained from the Dirac current. In Sec.~\ref{sec:three} we present an analysis based on momentum densities obtained from the canonical and from the symmetrized or Belinfante energy-momentum tensors, and show the conclusions are dramatically different from the Dirac current conclusions.  The twisted spinor fields always display a classical vortex behavior in the vicinity of the vortex line, and the vorticity is singular at the location of the line.  

These are conclusions reached from a field theoretic viewpoint, where there is a Lagrangian and a Noether procedure which, by studying the response of the system to coordinate translations,  gives the canonical energy-momentum tensor.  Certain components of the energy-momentum tensor give the momentum density, and the result is not the same as the above mentioned bilinear from the Dirac equation.  Further, the canonical energy-momentum tensor is not symmetric in its two indices.  This is a problem if one wants to use the energy-momentum tensor as a source in the General Relativity field equations, where symmetry is required.  One can symmetrize the energy-momentum tensor by adding a total derivative to the canonical result. The full momenta obtained by integrating components of the energy-momentum tensor are then the same in many circumstances, but the local momentum densities are not the same.  In the context of twisted electrons, while both the canonical and Belinfante cases give a singular vortex, the strength of the singularity is not the same.   Thus there are three local field momentum definitions to choose among, and the conclusions regarding the vorticity of the twisted electrons depends on the choice.  

Again, the plan of the paper is to display the twisted electron solutions in coordinate space in Sec.~\ref{sec:two} and make a beginning discussion of the vorticity from the Dirac current viewpoint.  In Sec.~\ref{sec:three} we will display the energy-momentum tensors derived from the Dirac Lagrangian and symmetrized by the Belinfante or Belinfante-Rosenfeld procedure, and show results for the vorticity in these cases. We will offer some closing commentary in Sec.~\ref{sec:four}.

%%%%%%%%%%%%%%%%%%%%%%%%%%%%%%%%%%

\section{Twisted electron wave functions and vorticity}

\label{sec:two}
%%%%%%%%%%%%%%%%%%%%%%%%%%%%%%%%%%

It has been known at least since~\cite{2011PhRvL.107q4802B} how to write relativistic twisted electron states, and this is also applicable to other spinor states.  What is newer is the discussion of the vorticity.  We will in this section review the twisted electron solutions, starting from a momentum space version (beginning where the component electron states have a common helicity), and discuss the vorticity properties of these states, obtaining in this section the velocity field from the Dirac current.  We will then write other solutions of interest~\cite{2017PhRvL.118k4801B,2017PhRvL.118k4802B} as linear combinations of these solutions and elucidate their vorticity properties.

A twisted electron state with vortex line passing through the origin is
\begin{align}
\ket{\kappa,j_z,k_z,\lambda} = A_0 \int \frac{d\phi_k}{2\pi} (-i)^\ell e^{i\ell\phi_k} 
\ket{\vec k, \lambda}		;
\end{align}
the state inside the integral is a momentum eigenstate with helicity $\lambda = \pm 1/2$ and momentum 
$\vec k = (k, \theta_k, \phi_k)$ in spherical coordinates. Longitudial momentum $k_z = k \cos\theta_k$ and transverse momentum magnitude
$\kappa = | \vec k_\perp | = k \sin\theta_k$ are the same for all states. Angle $\theta_k$ is the pitch angle.  The state normalization is
$\braket{\vec k',\lambda' | \vec k, \lambda}=
(2\pi)^3 2E \, \delta_{\lambda' \lambda}
\delta^3(\vec k' - \vec k)$, where 
$E= \sqrt{ k^2 + m^2}$.  The phases of the momentum eigenstates are in the Jacob-Wick convention~\cite{Jacob:1959at}.
These states are eigenstates of $J_z$ - the total angular momentum projected along the propagation direction - with eigenvalue $j_z$ and $\ell = j_z - \lambda$ must be an integer.  $A_0$ is a normalization constant.

The state in coordinate space is obtained using the Dirac field operator,
\begin{align}
&\psi_{\kappa,j_z,k_z,\lambda}(x)  = \braket{ 0 | \psi(x) | \kappa,j_z,k_z,\lambda }   \nn\\
&   \quad	= A_0 e^{i(k_z z - E t)}	\int \frac{d\phi_k}{2\pi} (-i)^\ell e^{i\ell\phi_k + i \vec k_\perp \cdot \vec\rho} \,
		u(\vec k,\lambda)    ,
\end{align}
where $u(\vec k,\lambda)$ is a Dirac helicity state.  With the traditional representation of the Dirac matrices~\cite{Bjorken:1965zz},
\begin{align}
u(\vec k,\lambda) = \left(	\begin{array}{c}
				\sqrt{E+m} \, \chi(\hat k,\lambda)		\\[1ex]
				 (2\lambda)	\sqrt{E-m}  \, 
				 \chi(\hat k,\lambda)		
				\end{array}		\right)	,
\end{align}
where
\begin{align}
\chi (\hat k, 1/2) &= \left(
\begin{array}{c}
\cos (\theta_k/2)  \\
e^{i\phi_k} \sin(\theta_k/2)
\end{array}              \right)   ,
                            \nn\\
\chi (\hat k, -1/2) &= \left(
\begin{array}{c}
-e^{-i\phi_k} \sin(\theta_k/2)   \\
 \cos (\theta_k/2)
\end{array}              \right)    .
\end{align}

Define~\cite{2017PhRvL.118k4801B}
\be
f_B^\ell = e^{i(k_z z - E t)+i \ell\phi_\rho} J_\ell(\kappa \rho)	,
\ee
where $\phi_\rho$ is an azimuthal angle in coordinate space.  The positive and negative helicity solutions are, first with $\ell = j_z - 1/2$,
\begin{align}
\psi_{\kappa,j_z,k_z,\frac{1}{2}}(x)  = \frac{A_0}{\sqrt{E+m}}
	\left(	\begin{array}{c}
		(E+m) \,  \cos (\theta_k/2)	\, f_B^\ell	\\[1ex]
		i (E+m) \,  \sin (\theta_k/2)	\, f_B^{\ell	+1}	\\[1ex]
		k \,  \cos (\theta_k/2)	\, f_B^\ell	\\[1ex]
		i k \,  \sin (\theta_k/2)	\, f_B^{\ell	+1}
				\end{array}		\right)
\end{align} 
and then, letting notation $\ell$ remain $\ell = j_z - 1/2$,
\begin{align}
\label{eq:negative}
\psi_{\kappa,j_z,k_z,-\frac{1}{2}}(x)  \!   =    \! \frac{A_0}{\sqrt{E+m}}  \
	\left(	\!\! \begin{array}{c}
		i (E+m) \,  \sin (\theta_k/2)	\, f_B^{\ell}	\\[1ex]
		(E+m) \,  \cos (\theta_k/2)	\, f_B^{\ell+1	}	\\[1ex]
		-i k \,  \sin (\theta_k/2)	\, f_B^{\ell}	\\[1ex]
		- k \,  \cos (\theta_k/2)	\, f_B^{\ell+1	}
				\end{array}		\!\! \right) .
\end{align} 
%The solutions as written have the same $\ell$ but different $j_z$.

The velocity field can be obtained from the Dirac current, $j^\mu = \bar \psi \gamma^\mu \psi$, thinking of $\vec j$ as density times velocity, so that
\begin{align}
\vec v = \frac{ \psi^\dagger \vec\alpha \psi}
    { \psi^\dagger \psi}    ,
\end{align}
and in the present representation
\begin{align}
    \vec \alpha = \left(
        \begin{array}{cc}
        0 & \vec\sigma \\
        \vec\sigma &  0
        \end{array}     \right) .
\end{align}
The vorticity, folowing standard definitions in fluid mechanics, is
\begin{align}
    \vec w = \text{curl } \vec v    ,
\end{align}
and $\text{curl } \vec v = \vec\nabla \times \vec v$ when the derivatives are well defined.  

Generically, examining the solutions for $\psi$ just given, the transverse components of the velocity field have the form
\begin{align}           \label{eq:vperp}
    \vec v_\perp = \frac
    { \ldots J_\ell(\kappa\rho) J_{\ell+1}(\kappa\rho)}
    { \ldots J^2_\ell(\kappa\rho) 
        + \ldots J^2_{\ell+1}(\kappa\rho) } 
    \hat\Phi
\end{align}
(or an equivalent with $\ell \to \ell -1$), where all the $\rho$ dependence is displayed and $\hat\phi$ is a unit vector in the $\phi_\rho$ direction.  For small $\rho$, $\rho \lesssim \ell/\kappa$, the Bessel functions are approximately
\begin{align}
    J_\ell(\kappa\rho) \approx \frac{1}{\ell!}
    \left( \frac{\kappa\rho}{2} \right)^\ell    .
\end{align}
We will take $\ell > 0$ for definiteness.  Then---usually---the $J_{\ell+1}^2$ term in the denominator can be neglected and 
\begin{align}
    \vec v_\perp \propto \rho \hat\phi.
\end{align}
This is not the velocity field of a classical whirlpool, but rather like the velocity field of water in a rotating bucket, with all locations of the water having a common angular speed.  The ($z$-component of) vorticity is constant at all locations where the linear $\rho$ dependence of $\vec v_\perp$ pertains, including at the vortex line itself (see, for example,~\cite{byron1992}).

However, the antiparallel or pure negative helicity $j_z = \ell - 1/2$ solution is special because the $J_\ell^2$ term in the denominator of Eq.~\eqref{eq:vperp} is suppressed in the paraxial (small $\theta_k$) limit.  One can see this by examining Eq.~\eqref{eq:negative}.  In full, the transverse velocity field in the antiparallel case is
\begin{align}
    v_\perp = \frac{\kappa}{E} 
\frac{ J_\ell(\kappa\rho)J_{\ell+1}(\kappa\rho) }
{ \sin^2(\theta_k/2) J_\ell^2(\kappa\rho)
+ \cos^2(\theta_k/2) J_{\ell+1}^2(\kappa\rho)  }
    \hat\phi        \,.
\end{align}
Now, at very small radii,
$\rho \lesssim \ell \tan\theta_k / \kappa$ (i.e., smaller by factor $\tan\theta_k$ than suggested above) ones still finds the ``water bucket'' like velocity rotation.  However, there is a significant region
\be         \label{eq:range}
    \frac{ \ell \tan\theta_k }{ \kappa}
    \lesssim \rho \lesssim 
    \frac{ \ell }{ \kappa}
\ee
where
\begin{align}
    \vec v_\perp \propto 
    \frac{ \hat\phi }{ \rho },
\end{align}
which is the velocity field of a classical whirlpool~\cite{poe1841}.  At any point with $\rho \ne 0$, the vorticity of this velocity field is zero~\cite{byron1992}.  

Were this $\rho$ dependence to persist to the vortex line, the vorticity at the vortex line would be singular, as one would show using the derivative free definition of the curl~\cite{byron1992},
\begin{align}
    (\text{curl } v)_z = \lim_{\sigma\to 0}
    \frac{1}{\sigma}    \oint \vec v \cdot \vec{dt} ,
\end{align}
where the integral is around the perimeter of a surface in the transverse plane, $\vec{dt}$ is a differential length tangent to the perimeter, and $\sigma$ is the area of the surface.  

We can now discuss the vorticity of the solutions in~\cite{2017PhRvL.118k4801B,2017PhRvL.118k4802B}.  (We should remark that~\cite{2017PhRvL.118k4801B,2017PhRvL.118k4802B} both limit the solutions at very large radii, to avoid the requirement of infinite energy for a pure Bessel wave, but this does not affect the low radius region needed for the vorticity discussion.)

Ref.~\cite{2017PhRvL.118k4801B} gives solutions with definite $j_z$.  A linear combinations of our existing solutions is
\begin{align}   \label{eq:bb}
\psi_{BB} &= \left( a \cos(\theta_k/2) -ib\sin(\theta_k/2) \right)
\psi_{\kappa,j_z,k_z,\frac{1}{2}}(x)
            \nn\\
&\hskip -2 em   + \left (-ia\sin(\theta_k/2) + b\cos(\theta_k/2)
\right)
\psi_{\kappa,j_z,k_z,-\frac{1}{2}}(x) ,
\end{align}
or
\begin{align}               \label{eq:BB2}
\psi_{BB} = \frac{A_0}{\sqrt{E+m}}
	\left(	\begin{array}{c}
		a (E+m)	\, f_B^\ell	\\[1ex]
		b (E+m)	\, f_B^{\ell+1}	\\[1ex]
		(a k_z -i b \kappa )	\, f_B^\ell	\\[1ex]
		(ia \kappa - b k_z )	\, f_B^{\ell	+1}
				\end{array}		\right)	.
\end{align}

Ref.~\cite{2017PhRvL.118k4801B} used a Weyl basis for the Dirac matrices.  Converting the above to its Weyl basis counterpart, it is easy to verify that for suitable choices of $a,b$, and $A_0$,  we get their first solution.  More specifically, choosing $b(E+m-k_z) - i a \kappa = 0$ will give zero second component in the Weyl representation, as they wish for this solution.  A similar procedure with $\ell \to \ell -1$ gets their other solution~\cite{bb2note}.

Their solutions may also be called parallel and antiparallel.  The parallel solutions shows no classical whirlpool behavior for any radius where the small argument expansion of the Bessel functions can be used.  The antiparallel solutions do have classical whirlpool solutions in the same range as Eq.~\eqref{eq:range}, but revert to the water bucket solutions for the smallest radii, so that neither solution has a singular vortex line.

%For the present discussion the important feature is that the Ref.~\cite{2017PhRvL.118k4801B} solutions will show the ``water bucket'' vorticity for all $\rho$ small enough that the monomial approximation to the Bessel function is useful, with no singularity.  

The Ref.~\cite{2017PhRvL.118k4802B} solutions can be obtained as a linear combination of a pure $b=0$ solution from Eq.~\eqref{eq:bb} plus a pure $a=0$ solution with $\ell \to \ell -1$.  This will be a linear combination of states with different $j_z$, but such is allowed.   The result is

%solutions are linear combinations
%\begin{align}           \label{eq:combo}
%&\psi_{B} = a  \left( \cos\frac{\theta_k}{2}  \ 
%\psi_{\kappa,j_z,k_z,\frac{1}{2}}
%-i\sin\frac{\theta_k}{2}  \ 
%\psi_{\kappa,j_z-1,p_z,-\frac{1}{2}} %\right)
%        \nn\\
%&\quad +
%b  \left( -i \sin\frac{\theta_k}{2}  \ 
%\psi_{\kappa,j_z-1,k_z,\frac{1}{2}}
%+ \cos\frac{\theta_k}{2}  \ 
%\psi_{\kappa,j_z,k_z,-\frac{1}{2}} %\right)  .
%\end{align}
%The combinations are chosen so that the upper components are proportional only to $f_B^\ell$, and one may also write
\begin{align}
&   \frac{ \sqrt{E+m}}{A_0} \  \psi_B =	                \nn\\
&(E+m) \left(	\begin{array}{c}
		a	\\[.3ex]
		b	\\[.3ex]
		0	\\[.3ex]
		0
		\end{array}	\right)	f_B^\ell
+	k_z \left(	\begin{array}{c}
	0	\\[.3ex]
	0	\\[.3ex]
	a	\\[.3ex]
	-b
	\end{array}	\right)	f_B^\ell
+	i\kappa \left(	\begin{array}{c}
	0	\\[.3ex]
	0	\\[.3ex]
	-b f_B^{\ell-1}	\\[.3ex]
	a  f_B^{\ell+1}
	\end{array}	\right)		.
\label{eq:fullbarnett}
\end{align}
This  is the Ref.~\cite{2017PhRvL.118k4802B} solution over all space, after inverting the Foldy-Wouthuysen transformation. Ref.~\cite{2017PhRvL.118k4802B} gives only the low transverse radius limit, which we can check upon expanding the Bessel functions.     
We omit the $f_B^{\ell+1}$ term on the grounds that it is generally small in the paraxial limit and particularly small in the very low $\rho$ limit that can be especially important.  We obtain
\begin{align}
&  \psi_B =	\, \frac{A_0}{ \sqrt{E+m}} \,  f_B^\ell 
    \, \times
        \nn\\
&\left\{  (E+m) \left(	\begin{array}{c}
		a	\\
		b	\\
		0	\\
		0
		\end{array}	\right)
+	k_z \left(	\begin{array}{c}
	0	\\
	0	\\
	a	\\
	-b
	\end{array}	\right)
-	i	\frac{2\ell}{ \rho e^{i\phi} }	\left(	\begin{array}{c}
	0	\\
	0	\\
	b	\\
	0
	\end{array}	\right)		\right\}		.		\label{eq:barnett}
\end{align}
This is the Ref.~\cite{2017PhRvL.118k4802B} solution for small $\rho$, verifiable upon noting (their notation) ${u_-}/{u_+}  = { k }/( E+m )$, which is the same as $k_z/( E+m )$ paraxially, and reworking the third term suitably.

For the vorticity analysis of these solutions, Eq.~\eqref{eq:barnett},
\begin{align}
    \psi^\dagger \vec\alpha_\perp \psi &=
    \frac{ 4\ell |A_0|^2 
        J^2_\ell(\kappa \rho) |b|^2}{\rho} \,
        \hat\phi    \,,     \nn\\
    \psi^\dagger \psi &=
    |A_0|^2     J^2_\ell(\kappa \rho)
    \bigg[ 2 E \left( |a|^2 + |b|^2 \right)
    + \frac{ 4 \ell^2 |b|^2}{ (E+m) \rho^2}  \nn\\
    &\hskip 5 em +\frac{ 4\ell k_z }{ (E+m) \rho }
    \Im \! \left( a^* b e^{-i\phi} \right)
    \bigg]  \,,
\end{align}
upon making a paraxial approximation.  Identifying behavior at moderate 
($ \ell  / E
\lesssim \rho \lesssim  \ell / (k \sin\theta_k) )$ 
and small ($\rho \lesssim \ell / E$) radii,
\begin{align}
    \vec v_\perp = \left\{
    \begin{array}{cl}
        \displaystyle \frac{ 2\ell }{E} 
        \frac{ |b|^2 }{ |a|^2 + |b|^2 }
        \frac{ \hat\phi }{\rho} & \qquad
            \text{moderate }\rho    ,    \\[2.5 ex]
        \displaystyle 
        \frac{E+m}{ \ell } \, \rho \, \hat\phi  & 
        \qquad  \text{small }\rho   .
    \end{array}         \right.
\end{align}

This is similar to the antiparallel single helicity solution.  But further, it is remarkable because the term that converts the classical whirlpool to the water bucket swirl is not important until a radius determined by the energy rather than the total momentum.  This is also true for the antiparallel spin solution of Ref.~\cite{2017PhRvL.118k4801B}, as can be worked out. This means that the transition radius is scaled by the Compton rather than the de Broglie wavelength, and the Compton wavelength is considerably smaller for nonrelativisitc and moderately relativistic situations.

%One can see the genesis of the result already in Eq.~\eqref{eq:combo}, where there is a paraxially small $b$ term that is crucial to the numerator of the velocity field expression, and also dominates the normalization in the denominator, but the latter only at small $\rho$.

The expression is a special linear combination, that satisfies the requirement that there be an important paraxially small contribution in order to obtain a classical whirlpool over a range of radii, but that cannot be hoped for generally, and at really small radii---at least for this method of calculating the velocity field---the non-singular water bucket like rotation will always pertain.    We have a physical example of a Rankine solution, which is an approximate solution known in fluid mechanics that has a transverse speed proportional to the inverse of the radius at large radii and proportional to the radius at small radii~\cite{acheson1990}.

%%%%%%%%%%%%%%%%%%%%%%%%%%%%%%%%%%

\section{Field theoretic vorticity calculations} 
\label{sec:three}
%%%%%%%%%%%%%%%%%%%%%%%%%%%%%%%%%%

For the Dirac field, starting from the Lagrangian one can derive an energy-momentum tensor, and from the energy-momentum tensor obtain the momentum densities of the field.  The straightforward or canonical procedure gives an energy-momentum tensor that is not symmetric in its two indices, and it can be symmetrized by a procedure due to Belinfante~\cite{1940Phy.....7..449B,rosenfeld1940energy} that does not change the total momentum obtained by integration, but does change the local momentum densities.   

The Dirac Langrangian density in its Hermitian form is 
$\bar \psi \big( (i/2)\overleftrightarrow{\slashed{\eth}} - m \big) \psi$.  
The momentum densities  that follow 
are
\be                 \label{eq:momdens}
\mathcal P^\mu = \left\{	\begin{array}{ll}
				\frac{i}{2} \bar\psi \gamma^0 \overleftrightarrow\eth^\mu \psi	,	&
					\quad\text{canonical},			\\
				\frac{i}{4} \bar\psi \left( \gamma^0 \overleftrightarrow\eth^\mu 
				+ \gamma^\mu \overleftrightarrow\eth^0 		\right) \psi		,	&
					\quad\text{Belinfante}	.
				\end{array}		\right.
\ee
%{\color{red} (Another note for us only: the canonical momentum density in the form above is Hermitian and is gotten by starting from the Hermitian form of the Dirac Lagrangian, 
%$\mathcal L = \bar\psi \big( (i/2) \overleftrightarrow {\slashed\eth}  -m \big) \psi$.   The Belinfante form, interestingly enough, comes out the same whether we start with the Hermitian Lagrangian or the more standard one.)           }
For the canonical momentum density and the Bessel solution, the radial momentum density is zero and
\be
\mathcal P^\phi_\text{can} = \frac{\ell}{\rho} \psi^\dagger  \psi	.
\ee
for all $\rho$.  We obtain the velocity field by taking the ratio of the spatial momentum density to the energy density. For either the canonical or Belinfante case,
\be
\mathcal P^0 = E \psi^\dagger \psi		,
\ee
so that,
\be
v^\phi_\text{can} = \frac{\ell}{\rho E}	.
\ee
This swirls at all radii,  and gives a singular vortex line at $\rho = 0$.  

This can also be interpreted as giving a momentum $p^\phi = \ell / \rho$ to a small test object that absorbs a twisted electron while located a distance $\rho$ from the axis.

Additionally, for the $z$-component,
\begin{align}
\mathcal P^z_\text{can} &= p_z \psi^\dagger \psi		,	\nn\\
v^z_\text{can} &= \frac{p_z}{E}		,
\end{align}
exactly as should be desired.

For the Belinfante case, visually the result looks (after the time derivative is turned into an energy) inbetween the canonical and Dirac results, and for the Bessel solution 
\be
v^\phi_\text{Bel} = \frac{ \ell}{2 \rho E}  
	+ \frac{ p_\perp }{2E} \frac{ J_{\ell+1}(p_\perp \rho)}{ J_\ell(p_\perp \rho)},
\ee
after some $p_\perp^2$ terms are dropped.  One also does get $v^z_\text{Bel} = {p_z}/{E}$,  with the dropping of some $p_\perp^2$ terms.  

These results show the canonical definition of the momentum density gives whirlpool-like swirling electrons and a definite vortex line with singular vorticity for twisted electrons, and the same is true for the Belinfante definition, but with only half the magnitude of vorticity at small radii.  

The field theoretical results can also be interpreted as giving a momentum $p^\phi = \ell / \rho$ to a small test object that absorbs a twisted electron while located a distance $\rho$ from the axis.  One would really like to know experimentally which calculation truly gives the momentum acquired by a such a test object.

%%%%%%%%%%%%%%%%%%%%%%%%%%%%%%%%%%

\section{Commentary}
\label{sec:four}
%%%%%%%%%%%%%%%%%%%%%%%%%%%%%%%%%%

We have discussed the vorticity of twisted electrons.  Twisted electrons are an example of structured electron wave fronts, for which, unlike in a plane wave, the velocity of the electron in the wavefront depends on precisely where it is.  There is then a velocity field, $\vec v(\vec x)$ (with possible time dependence tacit) and in analogy to fluid mechanics, a vorticity given by
\be
    \vec w = \text{curl } \vec v    .
\ee

One starting point for calculating the velocity is to use the Dirac current $\bar\psi \gamma^\mu \psi$, with the velocity obtained from
\be
    \vec v = \frac{ \bar\psi \vec \gamma \psi }
    { \bar \psi \gamma^0 \psi } .
\ee

Twisted photons have a swirling velocity, with a nominal orbital angular momentum along the direction of motion given by an integer $\ell$ or in general units by $\ell\hbar$.  The wavemfunction has a $\rho^\ell$ or $\rho^{\ell\pm 1}$, determined by the component, dependence, where $\rho$ is the transverse radius in cylindrical coordinates, approximately accurate up  to values of $\rho \approx \ell/\kappa$, where $\kappa$ is the transverse wave number of the component electrons that make up the twisted electron.  One has $\kappa = k \sin\theta_k$ where $k$ is the full wave number of the twisted photon components, and $\theta_k$ is usually a small angle.

We find, using wave functions available from~\cite{2017PhRvL.118k4801B,2017PhRvL.118k4802B} as well as wave functions made from electrons of definite helicity~\cite{2011PhRvL.107q4802B}, that the swirling within the radius given above is often well described as being like liquid in a rotating bucket, where all parts of the velocity field attain the same angular speed.  Such a velocity field has constant vorticty~\cite{byron1992} and no singularity at the vortex line.  Specially selected wave functions, particularly including but not limited to~\cite{2017PhRvL.118k4802B}, can have a classical whirlpool like velocity field, with transverse velocity $\propto 1/\rho$ in the region 
$\ell \tan\theta_k / \kappa
\lesssim \rho \lesssim  \ell / \kappa$.  But even for these solutions, the velocity field reverts to the water bucket distribution and no singularity for very small radii---for this way of calculating the velocity field.

This bring up what is perhaps a larger question, namely, what is the correct expression to use when calculating the velocity field?

In a field theory, starting from a Lagrangian, there follows by a canonical procedure an energy-momentum tensor, and selected components of this tensor give the momentum density and thence the velocity field.  The canonical energy-momentum tensor is not symmetric in its two indices, and this is a problem if using it as a source in the Einstein field equation general relativity.  The canonical energy momentum tensor can be symmetrized by adding a total derivative term, in a procedure worked out by Belinfante~\cite{1940Phy.....7..449B} and Rosenfeld~\cite{rosenfeld1940energy}, which does not affect calculations of total momentum but does change the local momentum density.   The canonical and Belinfante expressions for the momentum density were given in Eq.~\eqref{eq:momdens}.  The canonical expression is a gradient expression and gives a twisted electron velocity field with
\be
\vec v_{\perp,\text{can}} = \frac{\ell}{\rho E} \hat\phi
\ee
at all radii.  Thus there is a classical whirlpool like swirling of the electron field, with a singular vorticity at the vortex line.  The symmetrized or Belinfante result is also gives a classical whirlpool at small radii, with a vorticity about half as great as the canonical expression.  These results are quite different from the results that follow from the Dirac current.

There is an analogous situation, with a discussion (several examples are~\cite{1978OptCo..24..185H,2016NatPh..12..731A,Leader:2017htb,Afanasev:2022vgl}), for photons, where there are only two, the canonical and Belinfante, proposals for the momentum density, but also suggestions as to how one may determine which matches nature.  This is an important question which requires adjudication also in the electron case.

%%%%%%%%%%%%%%%%%%%%%%%%%%%%%%%%%%

\section*{Acknowledgements}

A.A.~thanks the National Science Foundation (USA) for support under grant PHY-2111063 and Army Research Office (USA) for support under grant W911NF-19-1-0022. C.E.C.~thanks the National Science Foundation (USA) for support under grant PHY-1812326.   A.~M. thanks the SERB-POWER Fellowship, Department of Science and Technology, Govt. of India  for support.

%%%%%%%%%%%%%%%%%%%%%%%%%%%%%%%%%%

\bibliography{vortelec}

\end{document}